\documentclass[12pt]{iopart}

\usepackage{graphicx,latexsym,color}
\begin{document}

\title{Towards a guided atom interferometer based on a superconducting atom chip}

\author{T M\"{u}ller$^{1,2}$, X Wu$^1$, A Mohan$^1$, A Eyvazov$^1$, Y Wu$^1$ and R~Dumke$^1$}

\address{$^1$ Division of Physics and Applied Physics, School of Physical and Mathematical Sciences, Nanyang Technological University, Singapore 637371, Singapore}
\address{$^2$ Physics Department, Faculty of Science, National University of Singapore, Singapore 117542, Singapore}
\ead{mueller@ntu.edu.sg}

\begin{abstract}
We evaluate the realization of a novel geometry of a guided atom
interferometer based on a high temperature superconducting
microstructure. The interferometer type structure is obtained with
a guiding potential realized by two current carrying superconducting
wires in combination with a closed superconducting loop sustaining a persistent current. We present the layout and realization of our
superconducting atom chip. By employing simulations we
discuss the critical parameters of the interferometer guide in particular near
the splitting regions of the matter waves. Based on measurements of the relevant chip
properties we discuss the application of a compact and reliable on-chip
atom interferometer.
\end{abstract}

\maketitle

\section{Introduction}
Superconductor technology combined with atom optical systems will allow a new generation of fundamental experiments and novel applications possibly reaching to the pairing of quantum solid state devices with neutral atoms on the quantum mechanical level \cite{Sorensen04,Andre06}. In particular, the enormous capability of superconducting devices paired with atom optics is especially promising utilizing microstructured surface traps. For example these can be used for quantum state transfer between solid state and atomic systems \cite{Sorensen04} or quantum information processing \cite{Andre06}. Microstructured surface trapping and
manipulation devices, so-called ''atom chips''
\cite{Fortagh07,Folman02}, have proven their great capability and
flexibility in the field of ultra-cold atom experiments
over the last years. These elements which allow the generation
of steep trapping potentials use current carrying
wires \cite{Fortagh98} or permanent magnetic structures \cite{Vuletic98} as well as micro-optical devices \cite{Birkl01}. They have for example been
used for guiding and transportation potentials of
various kinds \cite{Denschlag99,Dekker00,Hansel01} as well as for the generation of quantum-degenerate gases
\cite{reichel01,Aubin06}. Additionally, the manipulation of neutral atoms with microstructured elements is a promising approach for realizing systems suitable for quantum information processing \cite{Calarco00}. Recently, superconducting micro-structured elements have been successfully employed in atom-optics experiments \cite{Nirrengarten06,Mukai07,Roux08}.

A major application would be a guided chip-based atom interferometer: In comparison to
optical interferometers, atom interferometers have the potential
of being several orders of magnitude more sensitive for some
applications or giving access to classes of interferometric
measurements not possible with optical interferometry in
principle \cite{Berman97}. Because of the high intrinsic sensitivity, these interferometers
have to be built in a robust way to be applicable under a wide
range of environmental conditions.

A new approach to meet this challenge lies in the development of
miniaturized and integrated atom optical setups based on
micro-fabricated guiding structures. First experiments have been
carried out investigating chip based interferometer structures
\cite{Wang05,Schumm05,Shin05}. However, with the traditional atom
chips based on metallic conductors at room temperature a symmetric closed Mach-Zehnder interferometer type guiding
structure has not been realized. These structures based on normal conductors have always connections to external power sources impeding an ideally symmetric potential. However, the symmetry is one major ingredient
of guided atom interferometry. Until now, only radio-frequency
dressing \cite{Schumm05,Hofferberth06} or optical approaches \cite{Kreutzmann04}
realizing the beam-splitting as well as hybrid types of guided atom interferometers
which use magnetic confinement in combination with optical
beam-splitting pulses \cite{Wang05} can overcome this
limitation.

In this article we present the experimental design of a ceramic-based high-temperature superconductor (HTS) atom
interferometer chip structure using a persistent current loop to
overcome limitations which normal conductors are facing. In combination with two superconducting wires carrying an externally supplied current an atom guide is realized. Along the atomic propagation direction this guide is split into two arms. These arms separate the atoms further and are afterwards redirected to merge the atoms again. This realizes a structure similar to a Mach-Zehnder interferometer.

\begin{figure}[ht]
\begin{center}
\includegraphics*[width=7cm]{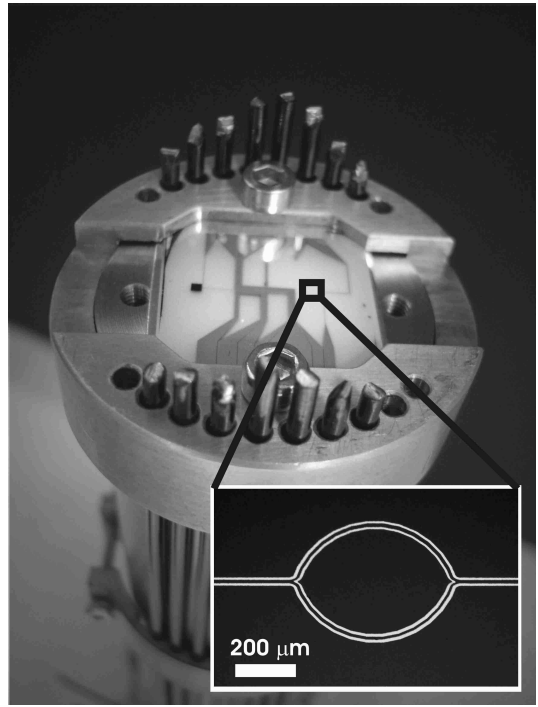}
\end{center}
\caption{Photograph of the microstructured superconducting atom
chip (size $20\times 20$mm$^2$) designed for atom trapping and manipulating, recorded before contacting the chip wires. The magnified
picture shows the interferometer structure recorded with an
optical microscope.} \label{fig:chipdesign}
\end{figure}

A major challenge of current carrying chip based atom
interferometers is the observed fragmentation of Bose-Einstein Condensates (BECs) close to the
wires \cite{Fortagh02,Leanhardt03}. Experimental investigations
show that this is caused by corrugations of the current carrying
wires \cite{Wang04}. This limitation has recently been suppressed
to some extent by optimization of the chip fabrication process
\cite{Schumm04} or with the use of alternating currents
\cite{Trebbia07}. With our epitaxially grown structure operated in the
superconducting state we expect a significant reduction of the
fragmentation effect.

These intriguing new experimental possibilities are the focus of
the work with our HTS atom chip shown in
figure~\ref{fig:chipdesign}. In section~\ref{sec:Atomchip} we
present the important material properties of the employed
superconductor. We compare different fabrication techniques for the
atom chip with regard to the quality and achievable size of
structures. Finally we discuss the expected fragmentation of
ultra-cold atoms for our atom chip based on the measured wire
corrugations and roughness. In
section~\ref{sec:interferometerguide} we discuss the design of the Mach-Zehnder type geometry by using two-wire
guides including a closed persistent current loop. We present numerical
simulations of the guiding potential based on a finite elements
method which are used for optimization of the potential. Finally,
we give an outlook for first experiments with the presented
atom chip and conclude in section~\ref{sec:conclusions}.

\section{Properties of the atom chip}\label{sec:Atomchip}

\subsection{Superconducting material properties}\label{subsec:superproperties}
The basis of the atom chip is a thin film of
YBa$_2$Cu$_3$O$_{7-x}$ (YBCO) on a Yttria-Stabilized Zirconia
(YSZ) single crystalline substrate. The thin
films used in our experiment typically have a critical temperature
T$_c$ of about 87K, depending on the exact composition of the YBCO. This
working temperature can be generated by using liquid nitrogen as
the coolant. This is a significant technical advantage compared to
non-HTS materials where the critical temperature can only be achieved with liquid helium. This means
the requirements for the cryogenic setup become much less
stringent by using HTSs which is an
important factor in the complex experimental setup of a typical
ultra cold atom experiment. In our case the critical current
density J$_{c}$ at liquid nitrogen temperature is 2~MA/cm$^2$.
For typical structure sizes on an atom chip with cross sections
of 100~$\mu m^2$ this results in an upper limit for the current of
2~A. This is a typical current value used for trapping in various
trap types on atom chips \cite{Fortagh07}.

Connected with the critical current density are the two critical
magnetic fields B$_{c1,c2}$ of the type-II superconductor.
Just like T$_{c}$, these parameters also crucially depend on the
exact composition of the YBCO. While B$_{c1}$ is typically on the order of
tens of mT, B$_{c2}$ can be as high as hundreds of T
\cite{nist}. Both these values support the application of YBCO in
atom chip structures where the employment of external magnetic
fields has to be taken into account. The externally generated
fields are usually on the order of a few mT, thereby reducing the
critical current density only by a minor amount. For
example, the external offset field in our experiment required for
the realization of a Ioffe Pritchard type trap with a Z-shaped
wire \cite{Denschlag99b} will be 6~mT. However, for future studies, such as suggested in \cite{Scheel07}, we can tune the superconducting
state from the Meissner to the Shubnikov phase by applying the
required offset magnetic field.

YBCO has been one of the most widely studied HTSs. To date, there are HTSs which possess even better properties than
YBCO concerning one or more of the relevant parameters discussed
above. However, the choice of the HTS is determined by the well developed micro-structure fabrication
techniques and structural properties of this thin film.

\subsection{Fabrication of the atom chip}\label{subsec:fabrication}
The manufacturing of the superconducting chip consists of two main
steps which are the fabrication of the HTS chip and its structuring. The YBCO is grown by epitaxy as a thin film on a
YSZ substrate. The lattice constants of YBCO and YSZ are matched,
allowing homogeneous growth of the superconducting material.
Another important factor is the matching of the temperature
expansion coefficient of the film and substrate. The atom chip
is subjected to stress by temperature changes of approximately 200-300K
in a short period of time ($<$5 min). A mismatch would result in
cracks of the superconducting film. The final thickness of the
YBCO film is 600-800nm, possessing a surface with a negligible
roughness determined by the epitaxic growth as discussed in
section \ref{subsec:fragmentation}. On top of the YBCO film a
200nm thick layer of gold is deposited. This film decreases the
electrical contact resistance. This is important when the
ceramic based superconductor is contacted to metallic electrodes supplying the
currents for the generation of the magnetic fields. Furthermore,
the gold film protects the superconducting material.

Structuring of the chip is performed by two different techniques.
These are standard optical lithography followed by a wet-chemical
etching as well as direct femtosecond (fs) laser ablation. The
standard lithography procedure has a resolution limit of about
$1\mu$m for the structure size. With the fs-laser ablation
procedure the desired patterns on the chip are realized by locally
removing the Gold and YBCO layer with focused laser pulses
(2$\mu$J, 130 fs, center wavelength 800nm), resulting in insulating
regions between the superconducting structures. In this laser
assisted structuring technique it is crucial that the film is not
heated. Due to heating oxygen would be lost and the
superconducting properties of the thin film would degrade. This
requirement sets the demand for using fs-laser pulses.

We compared the two methods of lithography and fs-laser machining by
measuring the roughness of the realized structures as will be discussed in
section~\ref{subsec:fragmentation}. We find that the structures machined
with standard lithographic techniques have a lower roughness than
the structures machined with the fs-laser. However, the laser ablation is
used in subsequently optimizing of imperfect regions of our
lithographically machined chips.

Additionally, we plan to employ e-beam lithography and ion-milling \cite{DellaPietra07} allowing for structure sizes less than $1\mu$m. These small sizes
will allow the fabrication of more complex structures and the
optimization of critical areas in the interferometer
structure.

\subsection{Roughness of the guiding structure}\label{subsec:fragmentation}

\begin{figure}[t]
\begin{center}
\includegraphics*[width=12cm]{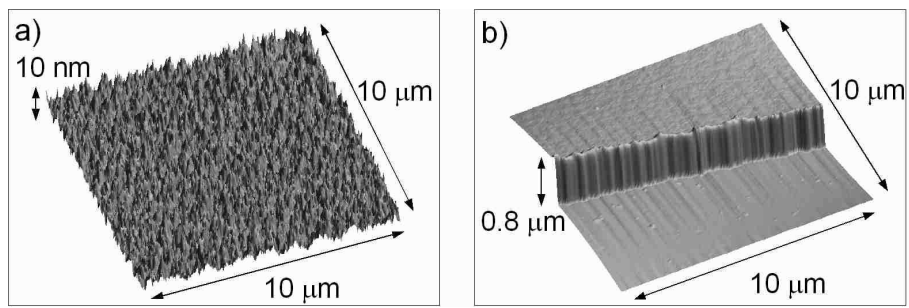}
\end{center}
\caption{Images of different areas with sizes of $100 \mu m^2$ of
the superconducting chip taken by atomic force microscopy (AFM). a) Scan above a flat surface of a
superconducting wire. b) Scan of a wire edge.
Note the different scales of the height amplitude. The straight
lines visible in part b) appear artificially in the AFM
measurement in the moving direction of the scanning tip induced by
the comparatively large height difference at the edge.}
\label{fig:AFM}
\end{figure}

The fragmentation of BECs and thermal atomic clouds trapped at
distances $\leq 100\mu$m from the current carrying wires has been
independently observed in several experiments on normal conducting
atom chips \cite{Fortagh02,Leanhardt03}. Since the atoms will be
held at a typical distance of 5$\mu m$ from the wires forming the
guide presented in this paper, these effects have to be taken into
account. The fragmentation is mainly caused by geometric
distortions of the current flow from the anticipated path in the
conductor, inducing an overall potential roughness \cite{Wang04}.
Corrugations of the wire edges have been identified as the main
contribution to these distortions \cite{Esteve04}. Besides the
employment of time-varying fields \cite{Trebbia07}, substantial
effort has been undertaken for optimizing the fabrication process
allowing for a suppression of the potential roughness. In this
respect, our atom chip meets the relevant criteria for a
successful implementation of optimal fabrication processes. The
chip structures are written by optical lithography \cite{Groth04} which has been
shown to reduce the roughness by two orders of magnitude
\cite{Kruger05,Kruger07}. To characterize the roughness of our
chip we performed measurements of the edges and surfaces of the
superconducting wires with an atomic force microscope, shown in
figure~\ref{fig:AFM}. The edges of the wires feature structures with typical sizes of about 200nm, which is
comparable to the grain size limit of 100nm measured in \cite{Kruger07}. Moreover, we measured the roughness
of the surface of the wire, which can also contribute to the
overall potential distortion. We find the roughness of the surface
almost negligible with a rms value of the surface height of only
2.3~nm. We explain this small roughness by one important intrinsic
advantage of YBCO, which is the epitaxic growth of the
superconducting film. The epitaxic growth of YBCO prevents basically all graining effects
during the fabrication. This graining has been identified to limit
the achievable conductor quality \cite{Kruger07}. We attribute the residual surface roughness to the gold film, possibly allowing
to achieve values below 1~nm for bare YBCO surfaces.

Beyond the known fabrication optimization procedure, our atom chip
promises further suppression of the fragmentation effect. As
pointed out in \cite{Kruger05,Kruger07}, for wires fabricated with
optimized procedures, inhomogeneities in the bulk material contribute to the potential roughness as well. A recent study demonstrated that scattering at inhomogeneities with different conductivity can become the dominating effect \cite{Aigner08}. Therefore we
anticipate a substantial reduction of this contribution due to the
resistance free current flow in the superconductor.

\section{Two-wire guide Mach-Zehnder interferometer structure}\label{sec:interferometerguide}
\begin{figure}[t]
\begin{center}
\includegraphics*[width=8.6cm]{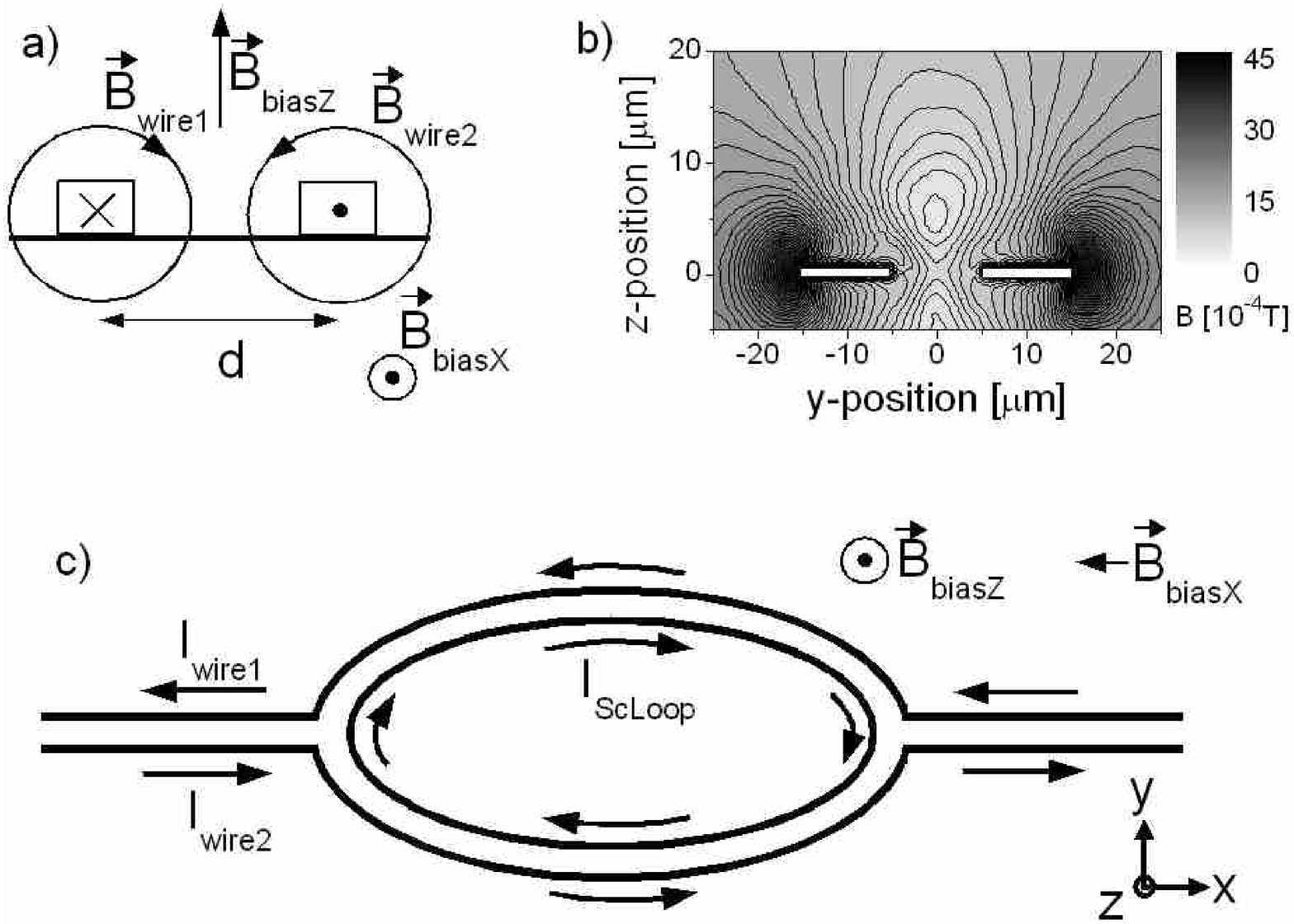}
\end{center}
\caption{Schematic of the two-wire guiding potential realizing a Mach-Zehnder interferometer-type geometry. The principle of the two-wire guide
using opposing currents and bias magnetic fields $B_{biasX}$ and $B_{biasZ}$ is
sketched in a). In b) the simulated magnetic field along the guide with wire currents of 55mA and
$B_{biasZ}=1.53$~mT, $B_{biasX}=0.3$~mT is shown. This basic guiding potential is transformed to
an interferometer type guiding structure by
enclosing a third wire in the form of a closed superconducting
loop as shown in c). This type of guiding potential is obtained
with opposing currents I$_{wire1}$ and I$_{wire2}$ and a persistent super
current I$_{ScLoop}$ of the same magnitude, which is always
directed opposite to the current in the neighboring wire.}
\label{fig:InterferometerSchematic}
\end{figure}

Figure \ref{fig:chipdesign} shows the pattern realized for the
chip structure. The heart of the chip is a lemon shaped closed
loop. This structure is designed so that a persistent current can
flow inside the loop. With this loop a novel two wire guide
geometry in a Mach-Zehnder-type interferometer structure has been
realized, schematically shown in
Figure~\ref{fig:InterferometerSchematic}. This unique design
ensures a symmetric splitting and symmetric recombination process
which is desirable for an interferometer type structure.

The guiding potential of this configuration is generated by two
parallel conductors carrying opposite currents and magnetic
offset fields perpendicular to the chip surface and along the wires, similar to \cite{Dekker00}. This results in a magnetic field minimum above
the chip surface (figure~\ref{fig:InterferometerSchematic}). In this
field minimum atoms can be guided. Our guide is realized by two
wires with widths of 10$\mu$m which are separated by a distance of
d=10$\mu$m. Typical values for our geometry are currents of 55mA
in each wire and bias magnetic fields of $B_{biasZ}=1.53$~mT and $B_{biasX}=0.3$~mT. This leads to oscillation frequencies for $^{87}$Rb in the $\left|F=2, m_F=2\right\rangle$ state of $\nu_y$=9.1~kHz in y- and $\nu_z$=8.2~kHz in
z-direction. This two wire guide can be extended to a
beam-splitter in the shape of a Y-junction by using a third wire
carrying a current always opposing the current of the neighboring
wire as suggested in \cite{Cassettari00}. The Y-junction formed with this two wire guide prevents
undesired loss channels. Additionally it minimizes changes in the guiding potential present in Y-junctions made of single-wire
guides. Unfortunately, with normal conducting wires in a planar
geometry the simple extension of this configuration into an
interferometer type topology made out of two opposing Y-junctions
is not possible, since the wires have to be connected to an external current
source. This drawback is overcome in our novel guiding design,
where the third wire placed between the two Y-junctions is formed
in a closed loop. By inducing a persistent current in this loop we can realize a Mach-Zehnder-type interferometer structure.

To investigate the properties of the interferometer structure we perform a numerical analysis by a finite elements method. The local current density in the wires is calculated which leads to the
magnetic field distribution. In combination with the additional
bias magnetic fields this results in the guiding confinement for the atoms. In this calculation we approximate the superconductor with effectively zero resistance and magnetic insulation. To verify these approximations we compare it to simulations based on the London equations and find no significant deviation in the guiding potential.

The structure of the guiding wires is optimized in order to
minimize changes in the guiding potential, in particular the transverse
oscillation frequencies near the separation point. These are the
main optimization criteria since a vanishing confinement at the separation point had been
identified to cause excitation into higher modes in a formerly
demonstrated beam-splitter on an atom chip leading to a random
phase in matter-wave interference \cite{Shin05}. For the optimization, we vary the values of the currents in the wires as well as the
offset magnetic bias fields for finding optimal guiding properties. Additionally, the
shape of the superconductor is optimized in order to decrease
the current density at sharp edges. The ratio of the current
density between the areas at the tip of the loop and the linear part is approximately two in the optimized geometry. Note that if the
local current density is reaching the critical value the
superconducting properties can be lost. After the separation
point, our structure has the shape of two separated two-wire
guides in the form of half-ellipses.

\begin{figure}[t]
\begin{center}
\includegraphics*[width=10cm]{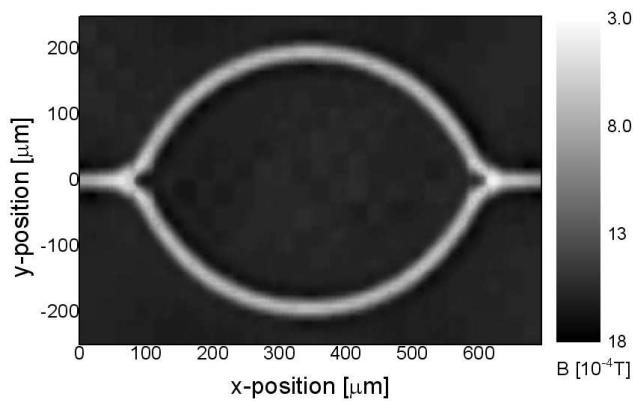}
\end{center}
\caption{Calculated absolute magnetic field in the x-y-plane at the height of the potential minimum in the bare two-wire guide.} \label{fig:Interferometer}
\end{figure}

\begin{figure}[t]
\begin{center}
\includegraphics*[width=10cm]{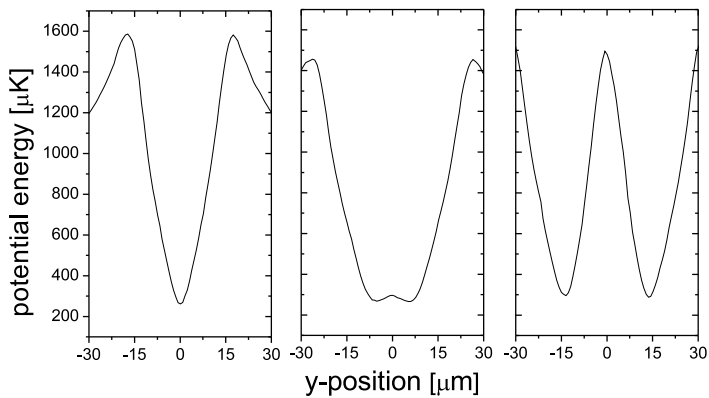}
\end{center}
\caption{Potential energy for $^{87}$Rb along the y-axis in splitting direction
for three different x-positions displayed as a temperature
$T=E/k_B$ for a height z=5.2$\mu m$ where $k_B$ is the Boltzmann constant. The x-positions correspond to a single guide, two guides shortly after the
separation point and two isolated atom guides, respectively.
(From left to right: x=47.3$\mu m$, 69.1$\mu m$, 81$\mu m$.)}
\label{fig:slices}
\end{figure}

\begin{figure}[t]
\begin{center}
\includegraphics*[width=10cm]{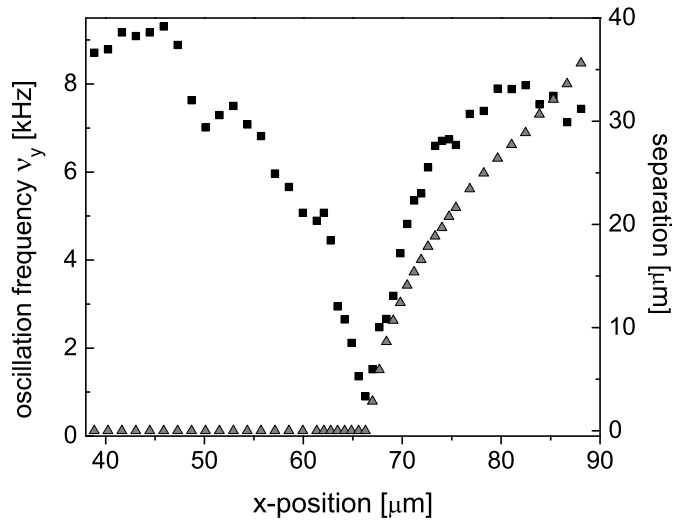}
\end{center}
\caption{Y-direction trapping frequency along the x-direction near
the separation point at the first Y-junction (boxes) and separation distance between the two potential minima of the guide (triangles). For x-positions
before the separation point at x=66.3$\mu m$ the calculation shows the
trapping frequency of a single atom guide, while after the
separation point the trapping frequencies in each of the two identical guiding arms are shown.}
\label{fig:trapfreq}
\end{figure}

The obtained numerical simulations of the
guiding structure for $^{87}$Rb atoms are shown in Figs.~\ref{fig:Interferometer} and
\ref{fig:slices}. Figure~\ref{fig:Interferometer} shows the absolute value of the
magnetic field in the x-y-plane above the surface of the wires at a height z=5.3$\mu$m. This corresponds to
the height of the potential minimum of the bare two-wire guide. In
figure~\ref{fig:slices} we show the potential in y-direction for
three chosen lines along the guide in x-direction at a height z=5.2~$\mu m$. The
calculated trapping frequencies in y-direction for different x-positions near the
separation point are shown in figure~\ref{fig:trapfreq}. Additionally we show in figure~\ref{fig:trapfreq} the distance between the two potential minima depending on the x-position after the separation. At the separation point the field minimum is at a height z=3.8$\mu$m and the oscillation
frequencies are $\nu_y$=0.9~kHz in y- and $\nu_z$=3.4~kHz in z-direction. Hence, we find strong
confinement and non-vanishing magnetic field at the separation
point. For longitudinal atomic velocities on the order of mm/s the adiabaticity of the splitting described by the quantity $(1/\nu^2_y)(\partial \nu_y/\partial x)(\partial x/\partial t)$ \cite{Shin05} can be ensured. Therefore we estimate the potential change not to be a limiting factor for the splitting process. After full
separation the potential of the guides in the two arms equals the
potential of the single two-wire guide. Due to the applied offset magnetic field we obtain a vanishing magnetic field at the merging point of the two arms. However, the influence of the magnetic field zero is not essential for the interferometer, since the atoms can be released from the guiding potential before reaching the absolute field minimum. After the separation point the minimum of the guiding potential is decreasing in propagation direction by 0.2mT with a maximum gradient which is one order of magnitude smaller than the gradient in confinement direction. Therefore, we do not expect excitation to higher transverse modes in the curved guiding structure.

To ensure the symmetry of the guiding potential, the current for the two outer wires will be applied in series, which also minimizes the relative current noise. To adjust the current in the closed loop we plan to employ spectroscopy of atoms brought close to the wires. A slight mismatch between the currents in the outer wires and the inner wire will not change the inherent symmetry of the interferometer structure.

The initialization of the atom guide and especially the persistent
current will be performed in a controlled temporal sequence by
changing the chip temperature and tuning the applied magnetic
fields. The persistent current in the closed loop can be controlled to
high precision by changing the offset magnetic field as presented in
\cite{Mukai07}. Additionally, in our setup we can tune the persistent current with the two half-loop wires of the
interferometer guide itself. Initially, the atoms will be prepared in a Z-shaped Ioffe-Pritchard type magnetic trap which is embedded in our superconducting atom chip (figure.~\ref{fig:chipdesign}). Subsequently, the atoms will be loaded into the wave guide by switching between different wire connections.

The stable operation of the interferometer type structure demands well-defined
control of the magnetic offset fields since field changes will
induce current changes in the superconducting loop. To reach 
a current stability of better than 0.01 in the closed loop (inductance L~$\approx$~0.3nH) the magnetic bias field $B_{BiasZ}$ has to have a stability of better than 1~$\mu$T. Therefore, this requirement demands a relative magnetic field stability of approximately $10^{-3}$ which can routinely be achieved in ultra-cold atoms experiments.

In first experiments with the guiding structure we will investigate the splitting
process of ultra-cold atoms at the Y-junctions, in particular its efficiency including undesired reflection and most
importantly its coherence properties. The
observations of ultra-cold atoms can be performed using standard
time-of-flight absorption imaging, allowing the overlapping of
split atomic clouds. For this the atoms can be released from the wave guide by switching off the current in the wires and the
superconducting loop. The persistent current in the closed loop can be changed by switching on an
appropriate magnetic offset field. The final operation of the interferometer type structure consists of
splitting, spatially diverging and finally recombining of ultra-cold atoms.

Our setup allows the evaluation of employing thermal atoms as well as quantum degenerate gases in the interferometer structure. By using a BEC the macroscopic matter wave characteristic is favorable, but may be limited by phase fluctuations in elongated condensates as measured in \cite{Hofferberth07,Jo07}. Alternatively, single particle interference can be employed with thermal atoms, where one can in principle investigate multi mode operation, similar to \cite{Andersson02}.

One major motivation in the field of guided
atom interferometry is their potential as compact and precise
inertial sensors \cite{Fortagh07}. In this respect, especially the
development of precision matter wave gyroscopes operated with
guided atoms is in the focus of current research due to the promising
realization of large enclosed areas for enhancing their
sensitivity \cite{Wu07}. Our current guiding potential encloses an area of
about 0.17~mm$^2$. An atom interferometer operated with
$^{87}$Rb enclosing this area already experiences a phase shift
for rotations of 33mrad/$\Omega_{earth}$ due to the Sagnac effect,
where $\Omega_{earth}=7.3\cdot10^{-5}$rad/s is the rotation velocity
of the earth. However, the enclosed area can in principle easily be scaled up, enhancing the sensitivity even further.

\begin{figure}[t]
\begin{center}
\includegraphics*[width=10cm]{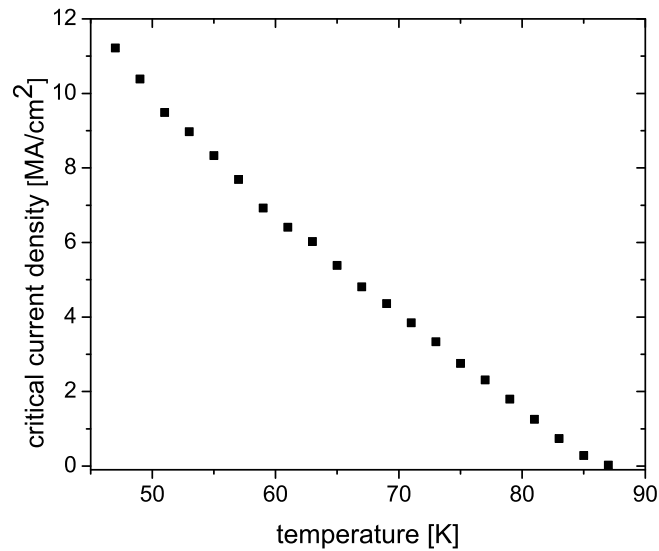}
\end{center}
\caption{Measurement of the temperature dependent critical current density of a 26~$\mu m$ wide wire. For lower temperatures the measurement is limited by the maximum obtainable current of 2A of the employed measurement system. The transition to superconductivity occurs at 87.1K.} \label{fig:currentdensity}
\end{figure}

In future experiments it might be interesting to operate the guide
at higher wire currents and with higher bias magnetic fields.
Consequently, tighter confinement can be achieved by employing
thin wires while maintaining a high current. In this respect,
superconductors offer an advantage compared to normal conductors since they
can provide current densities of more than $10^7$A/cm$^2$ without heat dissipation. To evaluate the limits of confinement of our
atom chip, we measured the critical current density at different
temperatures of the chip shown in figure~\ref{fig:currentdensity}. We find a
nearly linear increase in the critical current density for
decreasing temperatures up to the current limit of our measurement device. Our current setup allows the possibility to use liquid helium as a coolant as well. To estimate the critical current density at 4.2K, we conservatively follow the behavior given by $J_c(T)=J_c(0)\cdot\left[1-\left(T/T_c\right)^2\right]$. We find a critical current density of
13(1)MA/cm$^2$.

This current density allows to operate a 1$\mu m$ wide wire with a current of 60mA whereas the further reduction of the critical current density by accordingly increased bias magnetic fields has been taken into account. $^{87}$Rb atoms trapped with these parameters in a single wire guide at a distance of 1$\mu m$ would experience a trapping frequency of $\nu$=465~kHz. This trapping frequency corresponds to a Lamb-Dicke-parameter $\eta =\sqrt{\frac{\omega_{rec}}{\nu}}$ of 0.23 when interacting with laser light with a wave vector k=$2\pi/\lambda=2\pi/$780nm, where $\omega_{rec}=\hbar k^2/2m$ is the recoil frequency. This is an interesting regime for further applications such as for example quantum information operations \cite{Jaksch99}.

\section{Conclusions}\label{sec:conclusions}
In this paper we have presented a novel interferometer type guiding
structure based on a closed superconducting loop. The presented
guiding potential has been optimized with respect to critical parameters near the splitting regions of matter waves. The interferometer type guide has the potential to overcome limitations which current room temperature atom chip devices are facing. One example is the fragmentation of atomic clouds close to the chip surface. Currently, numerical simulations of the propagation of atomic matter waves in the presented guiding structures
are being performed for a further optimization of the second generation superconducting
chip devices. At present, the superconducting atom chip is
integrated into our cryogenic experimental setup for the generation of quantum-degenerate gases. The enormous potential offered by the combination of superconducting solid state systems with neutral atoms will allow a variety of novel experiments, which will not only be limited to high temperature superconductors. These are for example the probing of vortices in superconducting films with neutral atoms \cite{Scheel07}, the coupling of the magnetic moment of ultra-cold atoms to superconducting quantum interference devices \cite{Walraven81} and the quantum state transfer between solid state and atomic quantum systems \cite{Sorensen04,Andre06}. However, this type of experiments faces severe technical challenges which will partially be evaluated with the presented system.

\ack
We thank C.~H.~Oh for his kind support. Additionally, we thank L.~Wang
for assistance regarding the critical current density measurements. This work is
funded by the Nanyang Technological University (grant no. WBS M58110036) and by A-Star (grant no. SERC 072 101 0035 and WBS R-144-000-189-305).

\end{document}